\documentclass{sigchi}

 \toappear{ {\emph{MobileHCI '17}, September 04-07, 2017, Vienna, Austria \\
 \copyright~2017 Copyright is held by the owner/author(s). \\ACM ISBN 978-1-4503-5075-4/17/09. \\
 http://dx.doi.org/10.1145/3098279.3098529}
 }

\usepackage{balance}  
\usepackage{graphics} 

\usepackage[utf8]{inputenc}
\usepackage[ngerman]{babel}
\usepackage{txfonts}
\usepackage[pdftex]{hyperref}
\usepackage{color}
\usepackage{booktabs}
\usepackage{textcomp}
\usepackage{subfig}
\usepackage{enumitem}

\usepackage{microtype} 
\usepackage{ccicons}  

\def\plaintitle{Pharos: Improving Navigation Instructions on Smartwatches by Including Global Landmarks}
\def\plainauthor{Nina Wenig, Dirk Wenig, Steffen Ernst, Rainer Malaka, Brent Hecht, Johannes Schoening}
\def\plainkeywords{Global Landmarks; Landmark-based Navigation; Computer Vision; Smartwatches; Pedestrian Navigation}

\makeatletter
\def\url@leostyle{%
  \@ifundefined{selectfont}{
    \def\UrlFont{\sf}
  }{
    \def\UrlFont{\small\bf\ttfamily}
  }}
\makeatother
\def\pprw{8.5in}
\def\pprh{11in}

\definecolor{linkColor}{RGB}{6,125,233}
\hypersetup{%
  pdftitle={\plaintitle},
  pdfauthor={\plainauthor},
  pdfproducer={\plainauthor},
  pdfkeywords={\plainkeywords},
  bookmarksnumbered,
  pdfstartview={FitH},
  colorlinks,
  citecolor=black,
  filecolor=black,
  linkcolor=black,
  urlcolor=linkColor,
  breaklinks=true,
}

\begin{document}

\title{\plaintitle}
%

\numberofauthors{5}
\author{%
	\alignauthor{
	Nina Wenig, Dirk Wenig, Steffen~Ernst,~Rainer~Malaka\\
	    \affaddr{Digital Media Lab, TZI}\\
    	\affaddr{University of Bremen}\\
        \hspace*{-1em}\email{\mbox{$\{$nwenig,~{dwenig},~malaka$\}$@tzi.de} steffen.ernst@uni-bremen.de}}
    \alignauthor{
    	Brent Hecht\\
    	\affaddr{People, Space, and Algorithms (PSA) Computing Research Group}\\
    	\affaddr{Northwestern University}\\
   		\email{bhecht@northwestern.edu}}\\
    \alignauthor{
    	Johannes Sch\"{o}ning\\
	    \affaddr{Human-Computer Interaction}\\
	    \affaddr{University of Bremen}\\
    	\email{schoening@uni-bremen.de}}\\
}

\maketitle

\begin{abstract}
Landmark-based navigation systems have proven benefits relative to traditional turn-by-turn systems that use street names and distances. However, one obstacle to the implementation of landmark-based navigation systems is the complex challenge of selecting salient local landmarks at each decision point for each user. In this paper, we present \emph{Pharos}, a novel system that extends turn-by-turn navigation instructions using a single \emph{global landmark} (e.g. the Eiffel Tower, the Burj Khalifa, municipal TV towers) rather than multiple, hard-to-select local landmarks. We first show that our approach is feasible in a large number of cities around the world through the use of computer vision to select global landmarks. We then present the results of a study demonstrating that by including global landmarks in navigation instructions, users navigate more confidently and build a more accurate mental map of the navigated area than using turn-by-turn instructions.
\end{abstract}

\category{H.5.2.}{Information Interfaces and Presentation (e.g. HCI)}{User Interfaces --- \emph{input devices and strategies, interaction styles} }


%
%
\section{Introduction \& Motivation}
\label{sec:introduction}
Research across many fields has robustly established that landmarks are an essential means by which humans navigate through their environments~\cite{Foo2005Humans,Golledge1992Place,Golledge1999Wayfinding,Siegel1975Development}. People of all ages use landmarks in this fashion, and landmarks have been described as ``the key to the ability to orient oneself and to navigate in an environment.''~\cite{Sorrows1999Nature}. This literature has motivated researchers to develop a series of landmark-based navigation technologies that have been shown to outperform traditional turn-by-turn instructions in a number of studies, particularly in the case of pedestrian navigation~\cite{Goodman2005,Ross2004The,Tom2003Referring,Tom2004Language}.

\begin{figure}[t]
  \centering
  \includegraphics[width=\columnwidth]{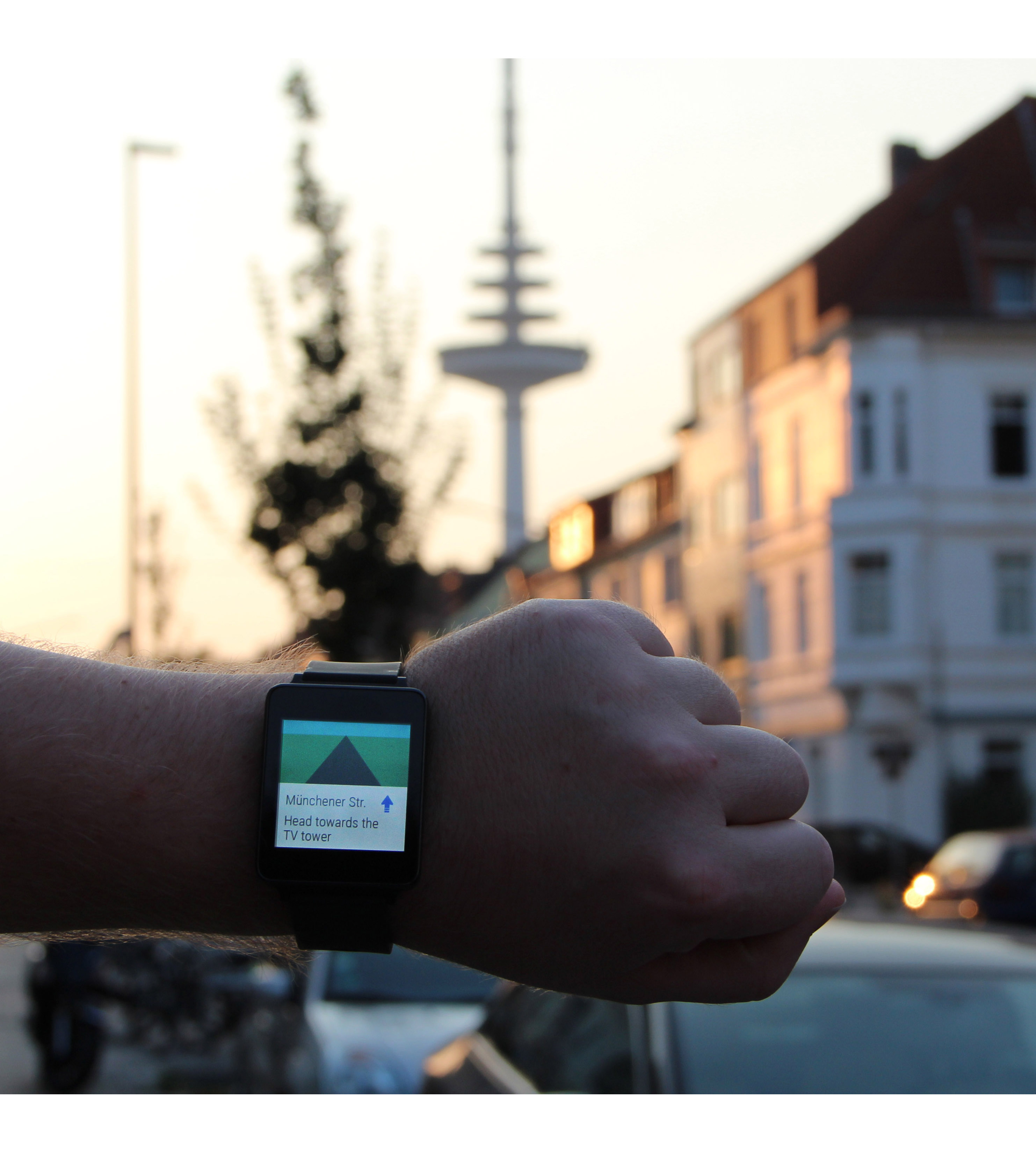}
  \caption{The Pharos navigation approach: Using Google Street View imagery, the visibility of global landmarks (here a municipal TV tower) is determined and then included in the turn-by-turn pedestrian navigation instructions for smartwatches.}
  \label{fig:pharos}
\end{figure}

However, despite the success of these research prototypes, well-known navigation technologies largely do not incorporate landmarks. This is in part due to the substantial implementation challenges associated with automating landmark-based navigation. For instance, landmarks that are salient for each individual user must be selected~\cite{Wakamiya2016,Winter2008Landmark}, the utility of specific landmarks for navigation is often gender- and language-dependent~\cite{Wakamiya2016}, and assessing the visibility of each landmark is difficult~\cite{Papadopoulos2011,Wither2013}.

In this paper, we introduce the \emph{Pharos} pedestrian navigation approach, which seeks to maintain the benefits of landmark-based navigation while substantially increasing the tractability of landmark-based navigation systems. The key to the Pharos approach is to reduce the challenges associated with landmark-based navigation by utilizing \textit{global landmarks}~\cite{Klippel2005Structural,Lovelace1999elements}, a class of landmarks that have not yet been considered in the navigation technology literature. The landmarks that are traditionally utilized are  \emph{local landmarks}~\cite{Winter2008Landmark} and are always located along route segments, particularly at key decision points. Global landmarks, on the other hand, are landmarks that can be seen from significant distances and can be located far from the user's route. Typical global landmarks are tall buildings (e.g. the Eiffel Tower, the Burj Khalifa, municipal TV towers), but mountains (e.g. Corcovado mountain in Rio de Janeiro, or the when there are mountain ranges in the background as in Denver, Seattle, Bishkek, Cape Town) or whole downtown areas (e.g. the skyline of New York) can also serve as global landmarks.

In addition to addressing the limitations of local landmarks discussed above (e.g. language/gender dependency), global landmarks also have several other important properties that are beneficial with respect to navigation tasks. First, global landmarks are constant, meaning that users can update their position and orientation relative to a single landmark rather than jumping from landmark to landmark. Secondly, because the distance to the landmark is not an important component of global landmark-based navigation, global landmarks can be used as a sort of compass for orientation~\cite{Steck2000Role}. Finally, while landmark-enriched navigation instructions require that the device's positioning system is able to locate the user with high accuracy for local landmarks, this is not true for global landmarks. A positioning error of a few meters might heavily influence the visibility of a local landmark  (e.g. a shop or a street sign), but the direction of a distant global landmark remains the same.

This paper demonstrates that the Pharos approach is both  (1)~feasible and (2)~has benefits compared to traditional turn-by-turn (i.e. no landmark) instructions. We establish feasibility by using computer vision (CV) and machine learning (ML) techniques to show that global landmark visibility is sufficiently extensive within cities and, critically, automatically detectable at scale with good accuracy. We also show that our approach works with publicly-available images and allows for pre-computation server-side, meaning that no further computation on a mobile or wearable device is needed.

We develop a prototype of the \emph{Pharos} approach and use it to establish the effectiveness of global landmark-based instructions through a field-based user study. Focusing on the use case of pedestrian navigation using smartwatches, we find that global landmark-enriched instructions outperform state-of-the-art turn-by-turn instructions along several key metrics. Specifically, we find that while both approaches had roughly similar performance in terms of navigation speed and number of errors, as hypothesized, Pharos outperformed turn-by-turn with respect to navigation confidence and the accuracy of the user's resultant mental map of the area.

To summarize, the contribution of this paper is four-fold:
 \begin{itemize}
   \item We introduce the Pharos navigation approach, which makes landmark-based navigation much more feasible by using global landmarks (relative to traditional landmark-based approaches that use local landmarks).
   \item We show that global landmarks are sufficiently broadly visible for pedestrian navigation tasks and that the locations at which global landmarks are visible can be computed at scale from geotagged imagery using computer vision and machine learning.	This information can be computed beforehand so that no processing on the wearable or mobile device is needed.
   \item We develop a pipeline to generate navigation instructions for pedestrians on smartwatches.
   \item We demonstrate through a user study that the Pharos global landmark-based navigation approach outperforms the current-state-of-the-art turn-by-turn instructions in important navigation metrics.
\end{itemize}

Below, we first introduce related work. We then describe the process by which we determine the visibility of global landmarks and create visibility maps. We next discuss the Pharos approach of integrating global landmarks into navigation instructions and demonstrate the benefits of our approach based on the results of a user study. Finally, we conclude with a discussion of the Pharos approach, landmark-based pedestrian navigation in general, and open problems in this area.

%
%
\section{Related Work}
This work draws from and builds on research in the domains of (1) pedestrian navigation and mobile guides, (2) landmark-based navigation in particular, and (3) landmark detection. Below, we discuss each of these areas in turn.

\subsection{Mobile Guides and Pedestrian Navigation}
Research on pedestrian navigation dates back almost two decades. From the beginning of this work, landmarks have played a role in guiding the user. For example, the GUIDE project~\cite{Cheverst2000Developing} aimed at integrating landmarks in textual navigation instructions. Malaka and Zipf~\cite{Malaka2000Deep} used a 3D city model to create navigation instructions not only incorporating the visibility but also the look of landmarks (e.g. ``turn right after the red building'') for pre-selected situations. Similarly, the LOL@ tourist guide for mobile devices~\cite{Pospischil2002Designing} enhanced routing information with references to landmarks. The PhotoMap \cite{Schoning:2009:PUS:1613858.1613876} uses images of taken with a GPS-enhanced mobile  phone as background maps for on-the-fly navigation tasks.

While one of the earliest mobile guides, the DeepMap system~\cite{Malaka2000Deep}, used a wrist-mounted display, wearable devices have become more relevant in research in the last five years. Indeed, in the context of pedestrian navigation, wearable devices do have an important benefit over their mobile counterparts as the user's hands can remain mostly free. Wenig et al.~\cite{Wenig2015StripeMaps} introduced StripeMaps, a cartographic approach for indoor navigation with smartwatches, which transforms 2D maps with route information into a 1D stripe. McGookin and Brewster~\cite{McGookin2013Investigating} investigated unidirectional navigation for runners by designing a navigation system called RunNav, which could also be used on a smartwatch. RunNav does not provide explicit routes, but rather a high-level overview to inform runners of areas that are good and bad places to run.

Even though research has explored how other modalities can be used to provide navigation instructions (e.g. auditory instructions~\cite{Holland2002} or haptic clues for mobile~\cite{Robinson2010} and wearable devices~\cite{Lim2015Vi-bros, Pfeiffer2015}), map-based navigation for mobile devices and turn-by-turn instructions for wearable devices have been established as de facto standards. For example, the current version of Google Maps for Android Wear smartwatches uses turns, street names, and distances to provide the user with instructions such as: ``After 20\,m turn left into Denver Road''.  Importantly, current navigation systems and applications typically do not include landmarks in their instructions.

\subsection{Landmark-based Navigation}
There is already a large corpus of related work exploring the benefits of landmark-based navigation for pedestrians. While the term \emph{landmark} was originally used to differentiate features with outstanding characteristics~\cite{Lynch1960Image}, the meaning of the term has changed over time and is now used more generically to describe well-known places~\cite{Golledge1992Place}. According to Sorrows and Hirtle~\cite{Sorrows1999Nature}, ``it is useful to understand landmarks in a way that supersedes knowledge in the environment''.

Landmark-based navigation instructions have been investigated in depth from the perspective of spatial cognition and cognitive psychology. Tom and Denis~\cite{Tom2003Referring} showed that for guiding pedestrians, route information referring to streets is less effective (regarding the number of stops, instruction checks and time) than route information referring to local landmarks. In another experiment by Tom and Denis~\cite{Tom2004Language}, the participants processed landmark-based instructions faster than street-based instructions and also remembered the route better with landmark information than with street information. Additionally, Ross et al.~\cite{Ross2004The} showed that adding landmarks to basic pedestrian navigation instructions (turn information and street names) results in less errors and a higher user confidence. In contrast to our work, neither Tom and Denis~\cite{Tom2003Referring,Tom2004Language} nor Ross et al.~\cite{Ross2004The} relied only on global landmarks in the navigation instructions.

From a more applied perspective, Wither at al.~\cite{Wither2013} showed that people can navigate solely using landmarks highlighted in panoramic imagery. Even `in the wild' in natural environments, as shown by Snowdon and Kray~\cite{Snowdon2009Exploring}, there are types of landmarks which are feasible to be used in mobile navigation systems. Recently, Bauer et al.~\cite{Bauer2016Indoor} have found evidence, that indoors pedestrian navigation instructions on mobile devices should only depict a single prominent landmark (instead of four) for high navigation efficiency. In contrast, the Pharos approach is primarily targeted at navigation in urban environments. While our work is related to the work of Wither at al.~\cite{Wither2013}, it is landmark-centric and instead of local landmarks the Pharos approach relies on global ones.

Landmarks do also play a role in designing pedestrian navigation aids for people with disabilities. For example, landmarks can be used to support low-vision and blind people in locating bus stops~\cite{Hara2013Improving} or to help mobility impaired users with navigating~\cite{Hara2016Design}. In addition, landmarks can be used for other purposes in location-based services. For example, Kray and Kortuem~\cite{Kray2004Interactive} interactively determined the user's positions based on the visibility of nearby landmarks. Lu et al.~\cite{Lu2010Photo2trip} generated traveling routes from geo-tagged photographs.

In general, landmarks have to be chosen carefully as their selection is highly important for the resulting quality of navigation instructions~\cite{Millonig2007Developing} (e.g. the personal salience of a landmark could be effected by language or gender~\cite{Wakamiya2016}). More specifically, Sorrows and Hirtle~\cite{Sorrows1999Nature} examined landmarks in real and electronic spaces and classified landmarks in terms of visual, cognitive and structural dimensions. Global landmarks have to be outstanding in all dimensions. Steck and Mallot~\cite{Steck2000Role} investigated the role of global and local landmarks in virtual environment navigation. They found out that in virtual environments both local and global landmarks are used for wayfinding, but different participants rely on different strategies; some of the participants used only local landmarks while others only used global ones. This results in a key question for our work, as with Pharos we aim at improving pedestrian navigation instructions with only a single global landmark rather than multiple local landmarks. 

\subsection{Landmark Detection}
Detecting landmarks is a crucial step in the pipeline of all location-based services relying on landmarks. While researchers have often opted to perform the selection of landmarks by hand~\cite{Malaka2000Deep,Millonig2007Developing} or using  crowd-based~\cite{Hara2013Improving} approaches, commercially successful applications require a solution that works on a global scale across a large set of users. As such, some automated approaches have been developed that make use of large databases of geographic features (e.g. OpenStreetMap [OSM]) \cite{Drager2012Generation,Raubal2002Enriching}. For example, Raubal and Winter~\cite{Raubal2002Enriching} explored landmark-enriched wayfinding instructions by automatically extracting local landmarks from a spatial dataset based on a formal measure for landmark saliency. More recently, Dr\"ager and Koller~\cite{Drager2012Generation} generated landmark-based navigation instructions for car navigation systems using OSM data. They used proximity to certain geographic features as a criterion to select landmarks. For prominent landmarks, the proximity can be used to estimate if a geotagged image shows a particular landmark~\cite{Papadopoulos2011}.

As global landmarks can be seen from longer distances and therefore the mere location information is not sufficient, visibility tests are always necessary. Visual landmark detection is related to the problems of location recognition, object recognition and image retrieval. For computer vision it is common to use simple image features, such as edges or corners in an image. For example, SIFT~\cite{Lowe1999}, SURF~\cite{Bay2006} and ORB~\cite{Rublee2011} are popular image feature descriptors, which can be used together with machine learning approaches, e.g. support vector machines~\cite{Chapelle1999} to predict whether an image contains a particular landmark. Most recently convolutional neural networks~\cite{Krizhevsky2012} have become very popular; they work directly on the pixel data.

Approaches for detecting different landmarks are often based on these computer vision algorithms in combination with the location of the landmark~\cite{Chen2011} or other contextual information~\cite{Li2009}. Zheng et al.~\cite{Zheng2009} built a large database from geotagged photos to detect the most popular landmarks around the world. As landmarks are visible from different positions, the visibility from different angles has to be computed~\cite{Hu2016alps}. Wither et al.~\cite{Wither2013} automatically created landmark-based navigation instructions by detecting salient landmarks in panoramic street imagery by using additional data, such as LiDAR information and text understanding. Such 3D information can often be used to enhance computer vision algorithms, e.g.~\cite{Hile2009Landmark}. Wakamiya et al.~\cite{Wakamiya2016} combined this information also with social data (Twitter or Foursquare) to determine the best local landmarks in an area.

%
%
\section{Pharos}
In this section, we describe the Pharos global landmark-based navigation approach. Pharos is named after the famous lighthouse in Alexandria that was one of the ``seven wonders'' of the ancient world and was used for centuries as a navigation landmark for ships. Below, we first report on a comparison of methods to determine the visibility of global landmarks. Secondly, we report on how we include global landmarks in the navigation instructions.

\subsection{Determining Visibility for Global Landmarks}
We determined the visibility of selected global landmarks to investigate whether they are (a) sufficient visible for pedestrian navigation tasks and (b) that this visibility can be detected at scale from geotagged imagery. The first step for this, is to compute their visibility for a given region of interest (ROI), e.g. the inner city or a district.

Global landmarks are diverse and range from mountains and large buildings to entire downtowns. The visibility of global landmarks is dependent on their \emph{prominence} in a ROI (rather than their absolute height). Prominence in a topographcial context describes ``the height of a mountain...summit by the vertical distance between it and the lowest contour line encircling it but containing no higher summit within it" \cite{noauthor_topographic_2017}.

Although global landmarks are often visible from large parts of a given ROI, they are not unconditionally visible from areas around the landmark as other buildings/structures might block line of sight. To create detailed visibility maps for global landmarks, we evaluated different computer vision and machine learning techniques to determine their visibility for a certain ROI.

\subsubsection{Techniques}
We compared the most common image features used in object detection tasks in combination with machine learning on three different landmarks: The Eiffel Tower in Paris (324\,m, France), the Petronas Towers in Kuala Lumpur (452\,m, Malaysia) and the Burj Khalifa in Dubai (828\,m, United Arab Emirates). All three are very prominent landmarks, but have very different characteristics (e.g. height, appearance, symmetry and structure). For each landmark, we manually collected as training images 150 (not necessarily geotagged) images containing the landmark using Google image search and Flickr. We made sure that the images for each landmark were taken in different lighting and weather conditions as well as from multiple perspectives and distances. In addition, as negative samples, we collected 450 images from Google Street View (GSV), which do not contain the landmarks but instead typical content in the ROI around the landmarks.

With this training data we compared SIFT~\cite{Lowe1999}, SURF~\cite{Bay2006} and ORB~\cite{Rublee2011} as image features in combination with a Support Vector Machine~\cite{Chapelle1999} and an approach using a convolutional neural network~\cite{Krizhevsky2012} (i.e. deep learning). For deep learning, it is common to use already-trained neural networks and refine them by training again on task-specific data to improve accuracy. Therefore, we used the convolutional neural network which was trained on the ImageNet data~\cite{Deng2009} and trained it in a second step with our dataset. For this, we used the DeCaf library~\cite{Donahue2014}.

We used a simple sliding window approach, in which we slide a window with the size $256{\times}256px$ in $64px$ steps over the image to create sub-images. Then, we tested for each sub-image whether it contains the landmark or not. We did this for different resolutions of the image by using an image pyramid to detect the landmark, even if the landmark is far away and very small or when it is larger than $256{\times}256px$ in the GSV image. We tested the sliding window against the selective search algorithm by Uijlings et al.~\cite{Uijlings2013} to find the landmarks in sub-images.

\subsubsection{Classifier Evaluation}
We used a separate data set with 100 manually collected GSV images for each landmark to evaluate the classifiers. We categorized an image as \emph{containing a particular landmark} when at least one sub-image (based on the sliding window or selective search approach) contains the landmark. The results of the comparison can be seen in Table~\ref{tab:f1-scores}; the $f1$-score is the harmonic mean of precision and recall. We calculated the scores across all three landmarks. The evaluation shows that we achieve the best results with the convolutional neural network combined with a simple sliding window approach. All results are similar for each landmark ($f1$-score Eiffel Tower: $77.3$, Petronas Towers: $79.5$, Burj Khalifa: $94.5$). The results for the Burj Khalifa are particularly good due to its prominence in Dubai's skyline. Overall, our results suggest that the visibility of global landmarks can robustly be detected at scale from geotagged images.

\begin{table}[tbp]
\centering
\begin{tabular}{lcc}
  & \textbf{Sliding Window} & \textbf{Selective Search} \\ \textbf{SURF+SVM} & 75.39 & 73.47 \\ \textbf{SIFT+SVM} &  69.26 & 60.70\\ \textbf{ORB+SVM} & 54.47 & 52.73 \\ \textbf{CNN} & \textbf{84.13} & 77.18
 \end{tabular}
\caption{Comparison of methods to determine the visibility of global landmarks. The $f1$-score is the harmonic mean of precision and recall. Our refined convolutional neural network outperforms the other approaches. In general, the sliding window approach works better than selective search.}
\label{tab:f1-scores}
\end{table}

\begin{figure*}[htbp]
\centering
\subfloat[]{
    \includegraphics[width=0.30\linewidth]{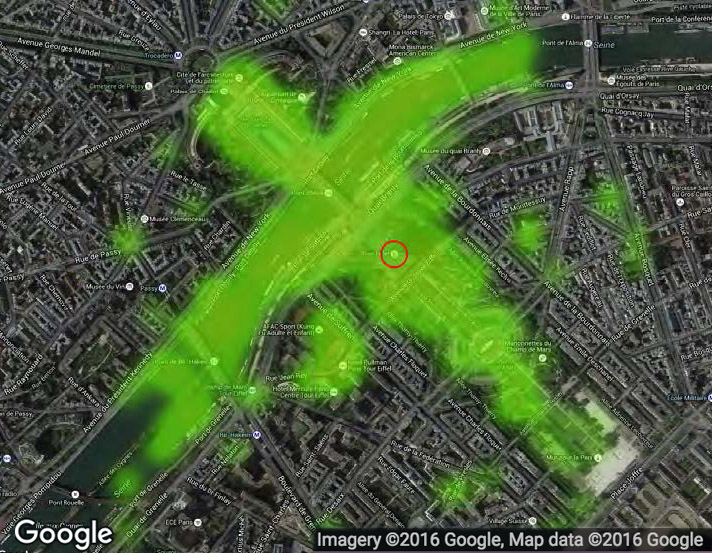}
    \label{fig:heatmap:eiffel}
  }
  \quad
\subfloat[]{
    \includegraphics[width=0.30\linewidth]{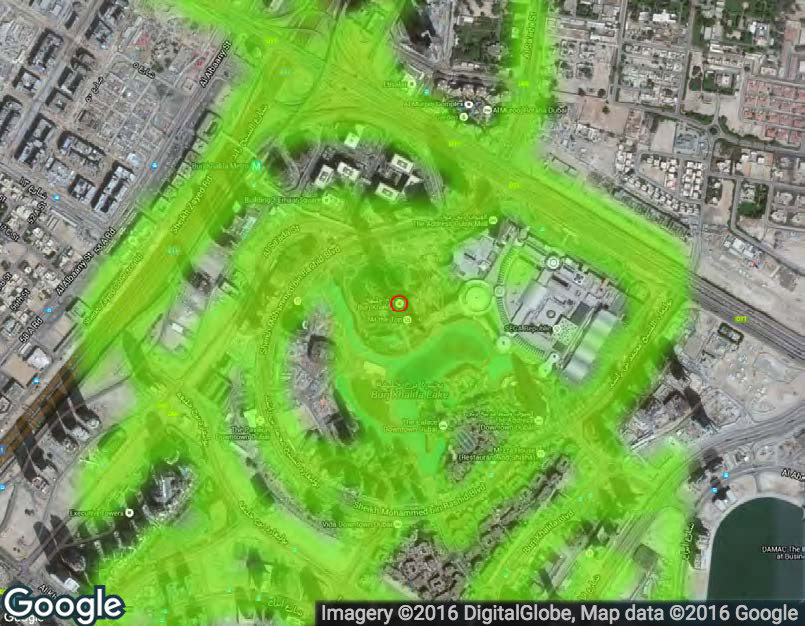}
    \label{fig:heatmap:burj}
  }
  \quad
\subfloat[]{
    \includegraphics[width=0.30\linewidth]{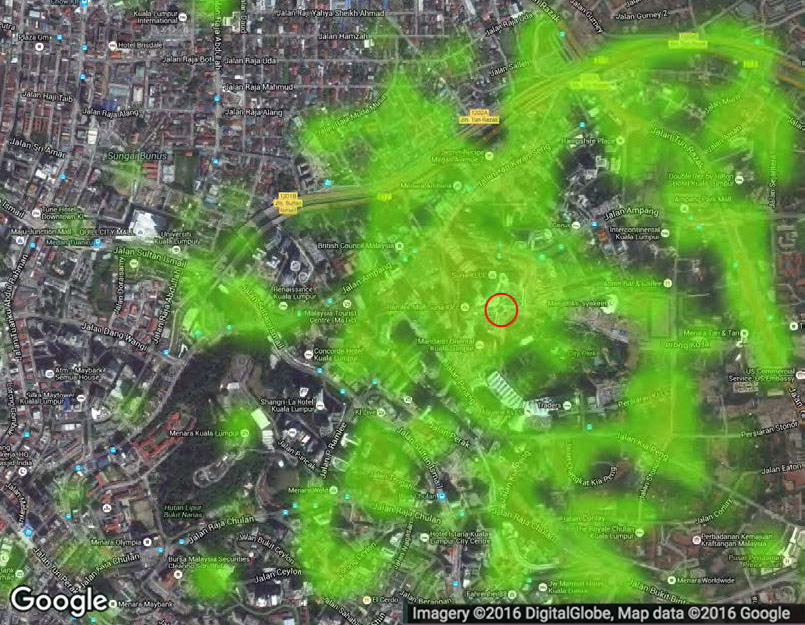}
    \label{fig:heatmap:petronas}
  }
  \caption{Visibility maps, automatically produced based on a CNN classifying Google Street View (GSV) images, for the three selected global landmarks: a) Eiffel Tower b) Burj Khalifa and c) Petronas Towers. Green spots indicate GSV images that contain the global landmark. Aerial images from Google Maps were used as base map (\copyright Google 2016).}
	\label{fig:visibilitymaps}
\end{figure*}

\subsubsection{Visibility of Global Landmarks}
As the results of the technical evaluation showed that the convolutional neural network works best, we used it to create visibility maps around the three landmarks using GSV images. In a radius of 2\,km around the landmark, we downloaded a GSV image at regular 100m intervals. This resulted in around 1600 images for every ROI. For some points or areas GSV is not available. After excluding these areas, we computed the visibility of the landmark in all the remaining images. Figure~\ref{fig:visibilitymaps} shows the resulting visibility maps for all three global landmarks. The green dots indicate that the landmark is visible in the GSV image. For the three selected global landmarks, the visibility maps show that these landmarks are likely sufficiently visible for pedestrian navigation tasks.

As the three selected global landmarks differ in shape, size and contour line of the surroundings, we conclude that determining the visibility of global landmarks based on public available imagery is feasible from the technical point of view. The visibility maps can be created on map and navigation servers beforehand so that no processing on the wearable or mobile device is needed. The visibility information than can be used to include global landmarks in navigation instructions.

\subsection{Including Global Landmarks in Smartwatch\\ Navigation Instructions}
Turn-by-turn navigation instructions inform the user about upcoming turns, usually combined with a notification (e.g. vibration). Global landmarks can be included in these instructions to confirm that the user has taken the correct turn and is still on the correct route. For Pharos, we developed a pipeline to include the visibility information of the global landmarks in turn-by-turn based pedestrian navigation instructions. While our approach could be applied to both car and pedestrian navigation systems, we opted to focus on the integration of global landmarks into instructions for pedestrian navigation. More specifically, we focused on smartwatch-based pedestrian navigation as smartwatches have important benefits relative to other mobile devices (e.g. smartphones) for navigation, e.g. they remove the need to constantly take a device out of one's purse or pocket~\cite{Wenig2015StripeMaps} (Note: we expect that our findings will generalize to a smartphone context, although further research is necessary to comform this hypothesis).

Current navigation systems for smartwatches, e.g. Google Maps for Android, use turn-by-turn-based instructions. They primarily rely on simple arrows showing the direction the user has to take at the next turning point (not at other decision points where the user does not have to turn), combined with the name of the street on which the user has to turn. Integrating global landmarks into such navigation instructions is not a trivial task. In general, situations in which navigation instructions can only rely on the global landmark are rare (e.g. ``Head towards the landmark'' when the route leads the user directly towards the landmark). Therefore, for Pharos, we aimed at enriching turn-by-turn navigation instruction with direct and indirect hints related to global landmarks. Direct hints precisely include the landmark in the navigation instruction, while indirect hints can be seen as additional information on the landmark's position relative to the user at decision points.

Textual instructions, especially on very small screens, need to be both short and understandable. For Pharos, we identified four different types of textual navigation instructions including a global landmark: the route heads towards towards or away from the global landmark, the global landmark is on the left or on the right, the global landmark is in a direction other than a cardinal direction (e.g. soft left), and the global landmark is not visible at the turning point.

The most simple situations are situations when the user, after a turn, walks straight towards the landmarks or walks away from the landmark. For such instructions, it is straightforward to add a direction to turn-by-turn instructions:

\begin{quote}
  \emph{``Head towards the landmark''} or\\\emph{``Head away from the landmark''}
\end{quote}

Similar are situations in which the landmark is on the left or on the right of the user after the turn, but cannot be directly included in the instruction. For such situations, we add indirect hints about the landmark's location, e.g.:

\begin{quote}
  \emph{``The landmark will be on your right''} or\\\emph{``The landmark will be on your left''}
\end{quote}

More difficult are situations in which the landmark is neither in front or behind the user nor on the left or on the right (e.g. 45 degrees or 135 degrees to the user's walking direction). Here, the English language struggles to unambiguously describe such situations without referring to angles. To address these situations, we use instructions such as:

\begin{quote}
  \emph{``The landmark will be in front of you to your right''} or\\\emph{``The landmark will be behind you on your left''}
\end{quote}

Additionally, there will be situations in which the landmark is not visible at the turning point. For these situations, we propose to include landmark information in the following form:

\begin{quote}
  \emph{``At the end of the street,\\
	the landmark will be in front of you''}
\end{quote}

From the technical point of view, all instructions can be easily generated using the visibility maps. For each turning point the angle between the direction of the following path segment (towards the next turning point) and the direction towards the landmark indicates the kind of instruction that should be used. For the last type of instruction, the visibility during the path segment also has to be considered. Whenever the landmark is not visible, simple turn-by-turn instruction can be used.

%
%
\section{User Study}
\label{sec:userstudy}
We performed a user study to evaluate the benefits of the Pharos approach and to explore how the inclusion of global landmarks changes the navigation experience. The study was focused on standard navigation evaluation metrics (time to reach destination, number of errors made) and the users' confidence that they were on the correct route. Confidence is not only an important aspect for the usability of a system in general~\cite{Brooke1996SUS}, it is particularly important for navigation systems (and its instructions) as the user is usually navigating in an unfamiliar environment with potential safety risks (e.g., other traffic participants). In addition, we also measured how well users built up spatial knowledge of the route using cognitive maps. Spatial knowledge supports users in performing the same or a similar navigation task without technological assistance the next time they are in the area and also aids them in identifying possible shortcuts~\cite{Golledge1992Place}.

\begin{figure}[htbp]
  \centering
  \subfloat[]{
    \includegraphics[width=0.4\columnwidth]{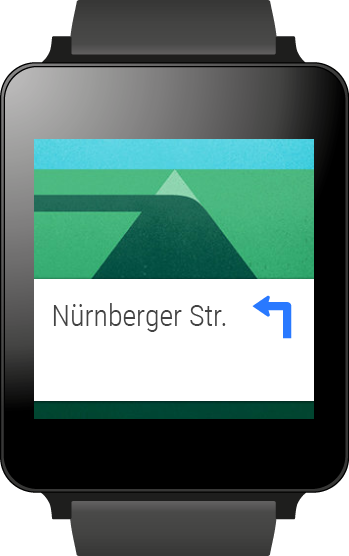}
    \label{fig:conditions:gm}
  }
  \qquad
  \subfloat[]{
    \includegraphics[width=0.4\columnwidth]{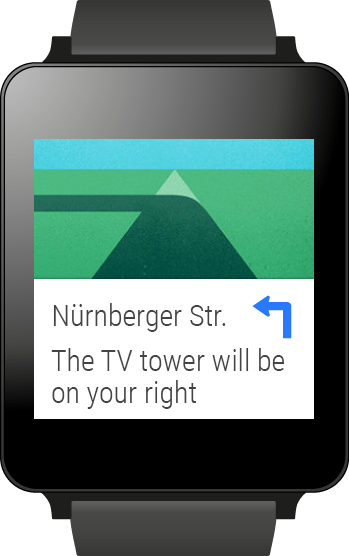}
    \label{fig:conditions:pha}
  }
  \caption{The two different conditions compared in the user study. Turn-by-turn navigation (TBT) on the left and the turn-by-turn navigation instructions including global landmarks following the Pharos approach on the right (PHA).}
  \label{fig:conditions}
\end{figure}

We compared the following two conditions in the user study, illustrated in Figure~\ref{fig:conditions}.

\begin{enumerate}[leftmargin=2em, label=\arabic*)]
	\item Turn-by-turn navigation instructions (TBT) as a baseline.
	\item Turn-by-turn navigation instructions enriched with global landmarks through the Pharos approach (PHA).
\end{enumerate}

In both conditions, the instructions are based heavily on Google Maps navigation instructions for Google Android Wear smartwatches. We used the same instructions as generated by Google Maps for Android Wear as well as the exact ``look and feel'' in the baseline (TBT) but enriched these instructions with the global landmarks in the PHA condition.

In the current version of Google Maps for Android Wear, the upcoming turn is shown with the remaining distance in meters (in steps of 10\,m). For the user study, we decided against this approach for two reasons. First, in pre-tests, the distance measures and the position of the notifications were insufficienctly accurate as they occurred often with an offset of around 10\,m. Second, pedestrian navigation instructions should neither require nor encourage the users to constantly check the system, distracting them from their primary task of walking.

To achieve optimal comparability, in the baseline condition (TBT) as well in the PHA condition we followed a wizard-of-oz study approach with instructions manually triggered by an experimenter. Participants were notified of new instructions via vibration of the smartwatch.

\subsection{Participants \& Apparatus}
The study was conducted in a residential district of Bremen (Germany), a mid-sized city in northern Europe. As a global landmark, we used a telecommunication and television tower (referred as TV tower in the rest of the paper), which is about 235 meters high. It can be seen from significant distances (see \nameref{sec:introduction}). The ROI of the study features wide and narrow streets with mid-sized row houses (up to three or four stories) such that the global landmark is not visible at all times.

\begin{figure*}[tb]
  \centering
  \subfloat[]{
    \includegraphics[width=\columnwidth]{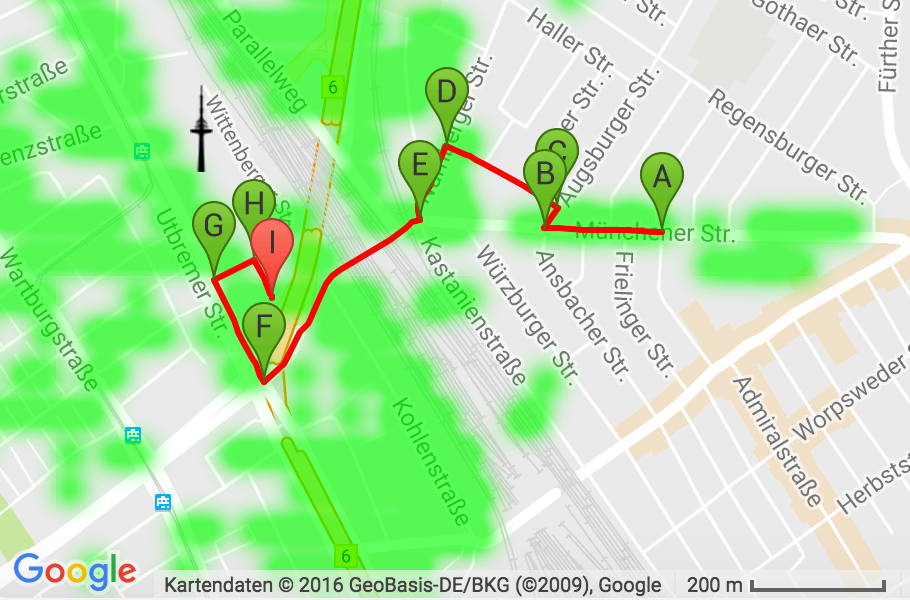}
    \label{fig:routes:r1}
  }
  \quad
\subfloat[]{
    \includegraphics[width=\columnwidth]{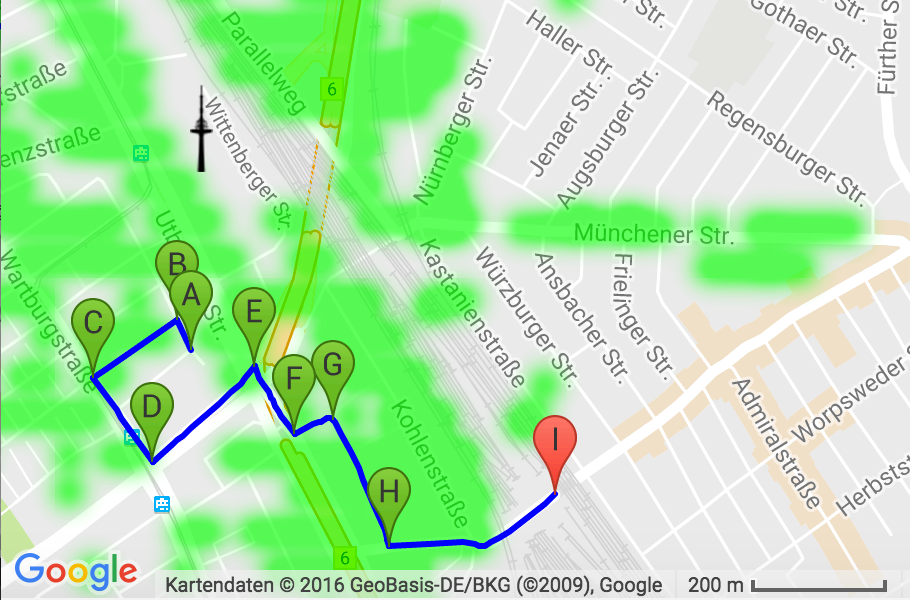}
    \label{fig:routes:r2}
  }
  \caption{The two routes used in the user study. Green dots indicate the visibility of the TV tower based on GSV images. Route 1 (a, red) has a length of 1.12\,km and route 2 (b, blue) is 1.17\,km long. For both routes, \emph{A} is the starting point and \emph{I} is the end point. Google Maps was used as base map (\copyright Google 2016).}
  \label{fig:routes}
\end{figure*}

For the user study, we selected two routes. Each of the routes had a very similar lengths (route 1: 1.12\,km, route 2: 1.17\,km) and featured an equal number of turns (seven turns plus the start and end point). The maximal length between two turns on both routes was 0.3\,km. The global landmark was similarly visible in both routes. Figure~\ref{fig:routes} shows both routes on the visibility map for the TV tower.

The turn-by-turn-based navigation instructions were created as follows. First, we used the multi-destination feature of Google Maps to create the route. Secondly, we followed the route using Google Maps for Android Wear and at all turning points we took a screenshot of the smartwatch (including the background image showing the turn in addition to the arrow, see Figure~\ref{fig:conditions}). We used these screenshots to create the instructions and (for both conditions) removed the countdown distance measures (see above). For the Pharos condition, we enriched the instructions with information about the location of the TV tower (as described in the previous section). At one of the turning points and the following route segment of route~1, the landmark was not visible at all. At this point we used the baseline TBT instruction also in the PHA condition.

The instructions for the next turning point were shown immediately before the turn at the exact same locations (manually triggered by the experimenter accompanied with a vibration notification of the new instruction). As a result, on the next route segment (after the turn) the instruction would be outdated. To exclude irritations because of outdated instructions, for the path segments between the turning points we slightly adapted the navigation instructions for both conditions. These navigation instructions were based on the instructions for the previous turning point, but the turning arrow was replaced with a straight ahead arrow. Furthermore, we changed the tense of the textual instruction from future tense to present tense (e.g. ``The TV tower is on your right'' instead of ``The TV tower will be on your right''). In contrast to the turning point instructions, these in-between instructions were not accompanied by vibration.

In accordance with our wizard-of-oz study design, we built a simple Android Wear smartwatch app to present the navigation instructions. An Android companion app we built for smartphones (showing all the turn-by-turn-based navigation instructions in a list view) allowed the experimenter to select and send the instructions to the smartwatch. The experimenter triggered the instructions always at the same predefined spots, approximately 5\,m before the turning point. Furthermore, the app allowed the experimenter to count how often the participants looked at the smartwatch.

We recruited 12 participants (4 females, 8 males) with an average age of 27.1 ($SD{=}2.5$). Most of our participants had very limited familiarity with the study area, although some had traveled along minor segments of the routes before. All participants own a smartphone and two of the participants regularly wear a smartwatch, but only one of them had navigated via smartwatch before. Ten of the participants use a smartphone for navigation purposes on a regular basis.

The user study was conducted with a LG G Watch. All participants performed the test in both conditions (within-subject design). The orders of the two conditions (TBT and PHA) as well as the two routes (route 1 and route 2) were counterbalanced.

\subsection{Task \& Procedure}
The participants were introduced to the experiment and told to follow the two different routes, one after the other. We explained the navigation task but did not mention the role of the TV tower. The participants did not have to select a route or target destination on a mobile device or on the smartwatch. For both conditions, we oriented the participants in the direction of the first movement and then started the navigation task with the first instruction.

As participants walked the route, the experimenter followed a few meters behind, selecting the navigation instructions on the companion app, collecting timing information, counting the number of times the user looked at her watch, and assessing the number of navigation errors. An error was assessed when a participant took a wrong turn without noticing their mistake after 10 meters (after this point, the experimenter would then guide the participant back on the route). While they were navigating, we asked the participants to rate their confidence (i.e. whether or not they believed they were on the correct route) on a seven-point scale. We performed this confidence assessment three times: after the start instruction, in the middle of the route, and at the end of the route.

\begin{figure*}[tbp]
  \centering
  \subfloat[]{
    \includegraphics[width=0.30\linewidth]{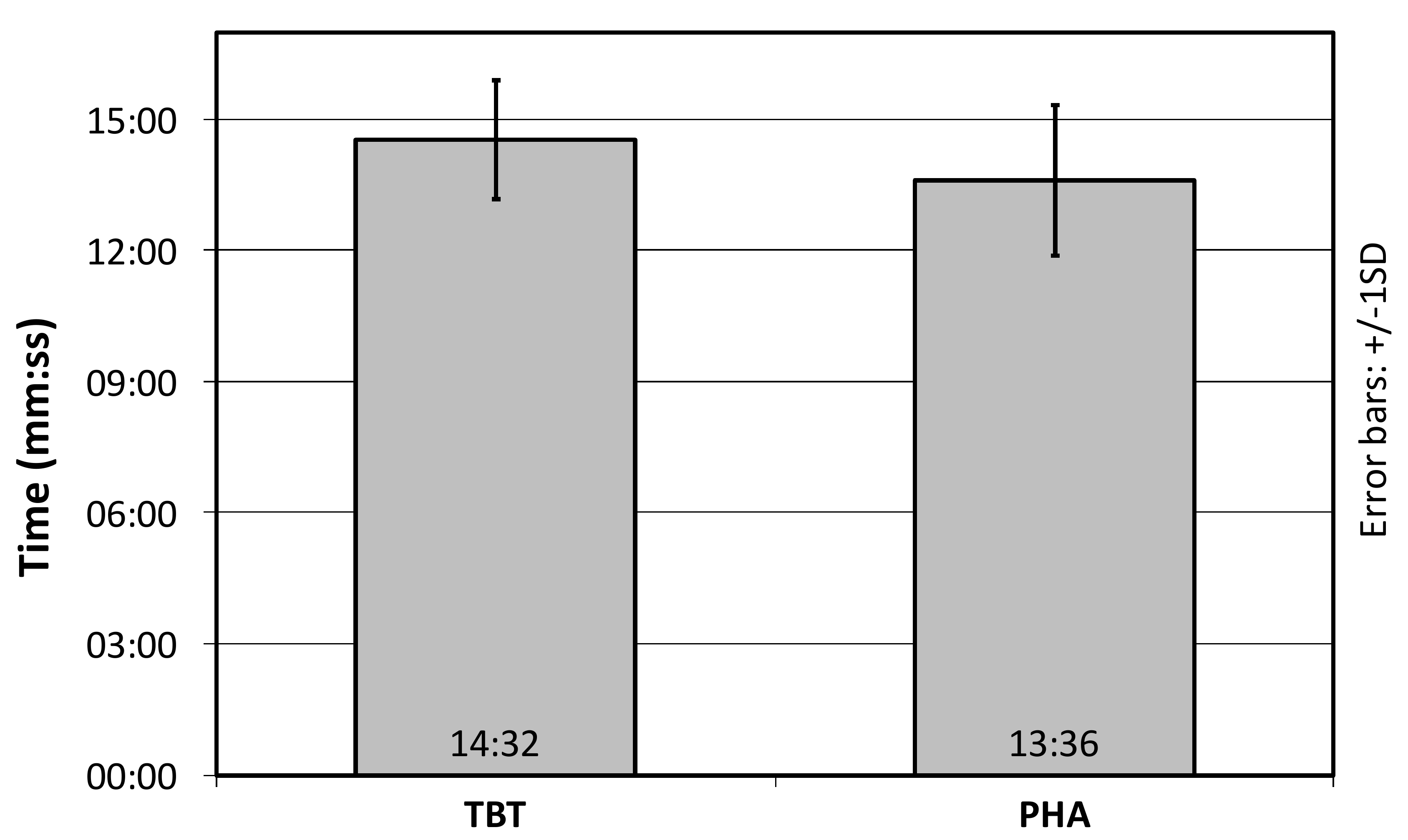}
    \label{fig:results:time}
  }
    \quad
  \subfloat[]{
    \includegraphics[width=0.30\linewidth]{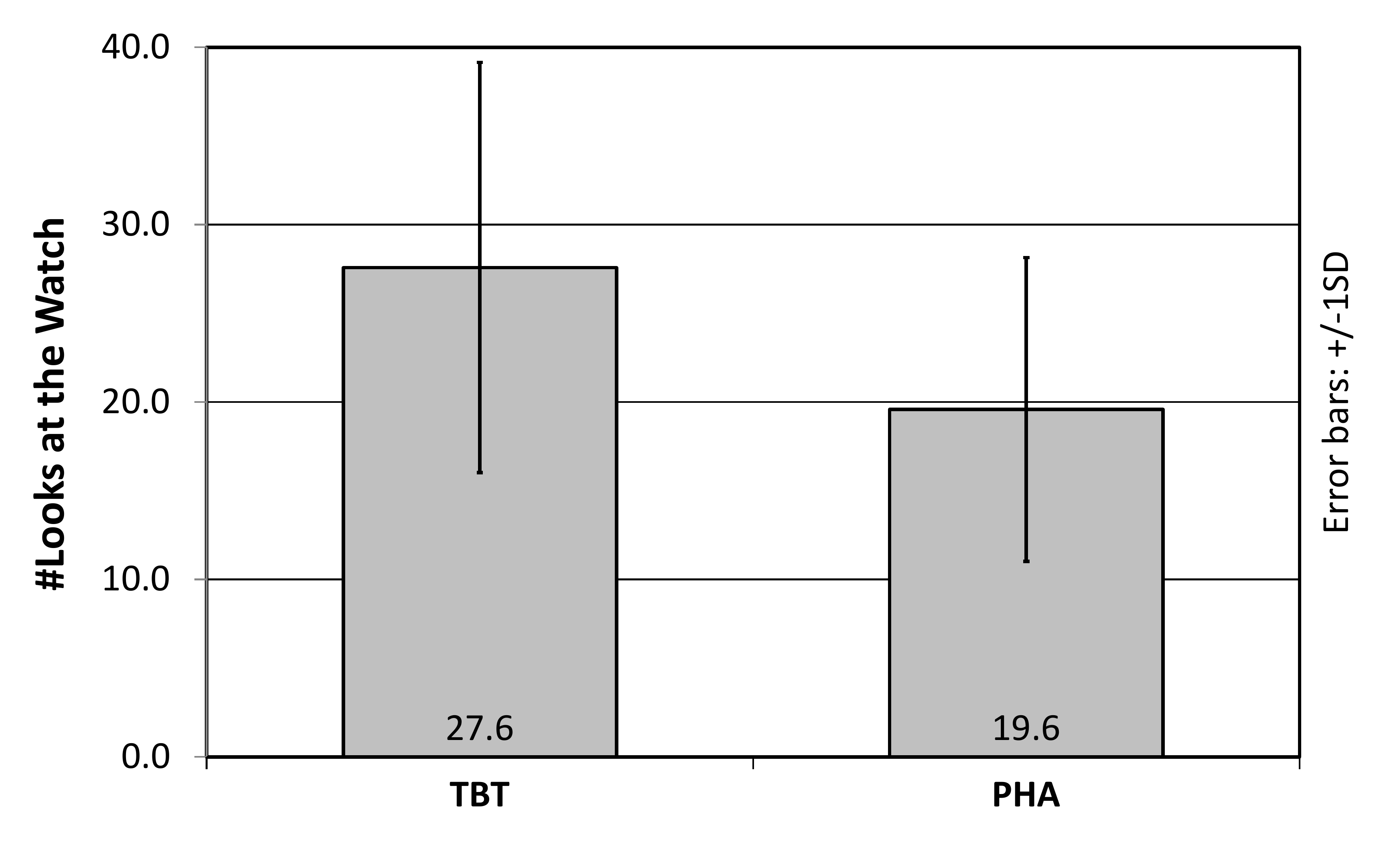}
    \label{fig:results:looks}
  }
  \quad
 \subfloat[]{
    \includegraphics[width=0.30\linewidth]{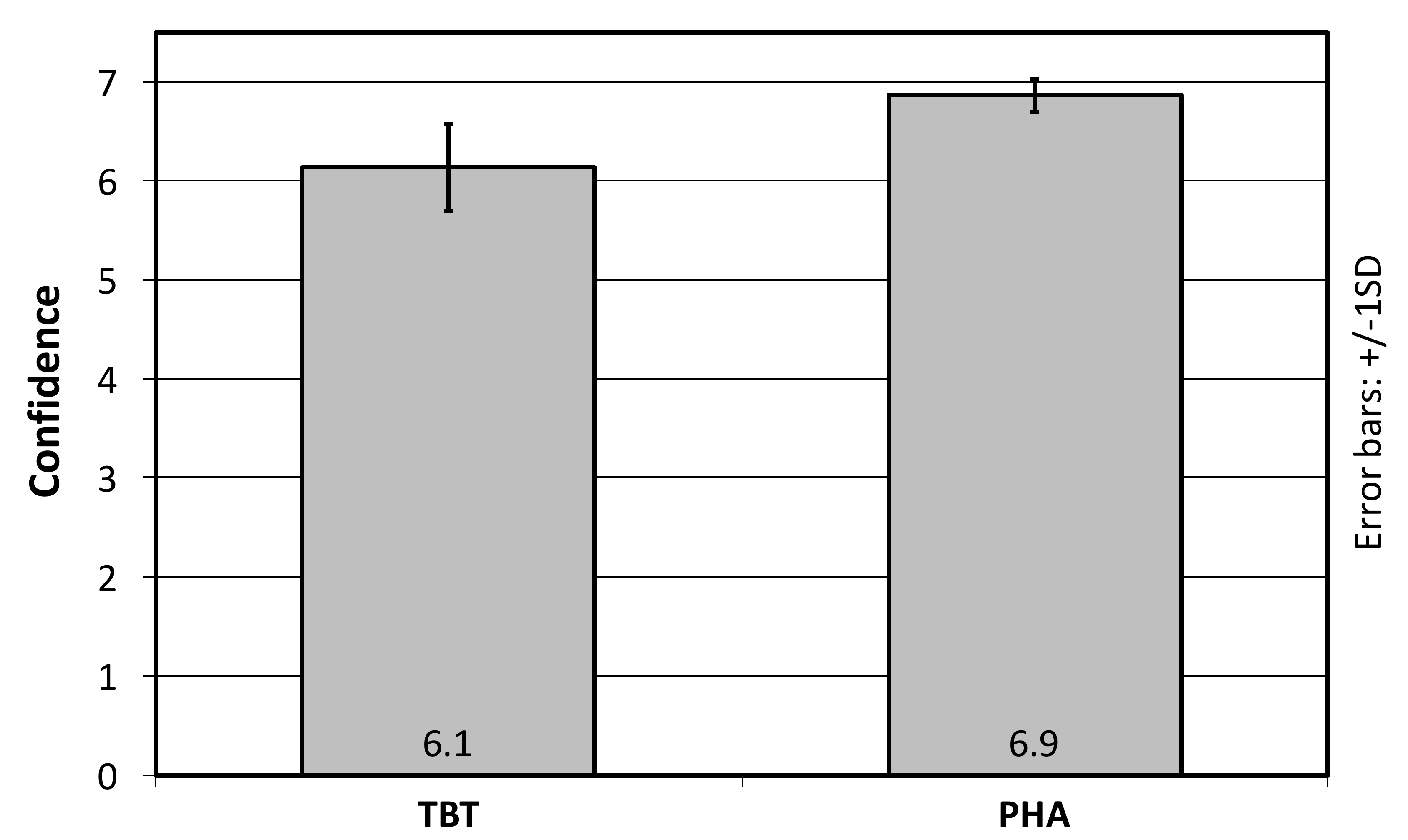}
    \label{fig:results:conf}
  }

  \subfloat[]{
    \includegraphics[width=0.30\linewidth]{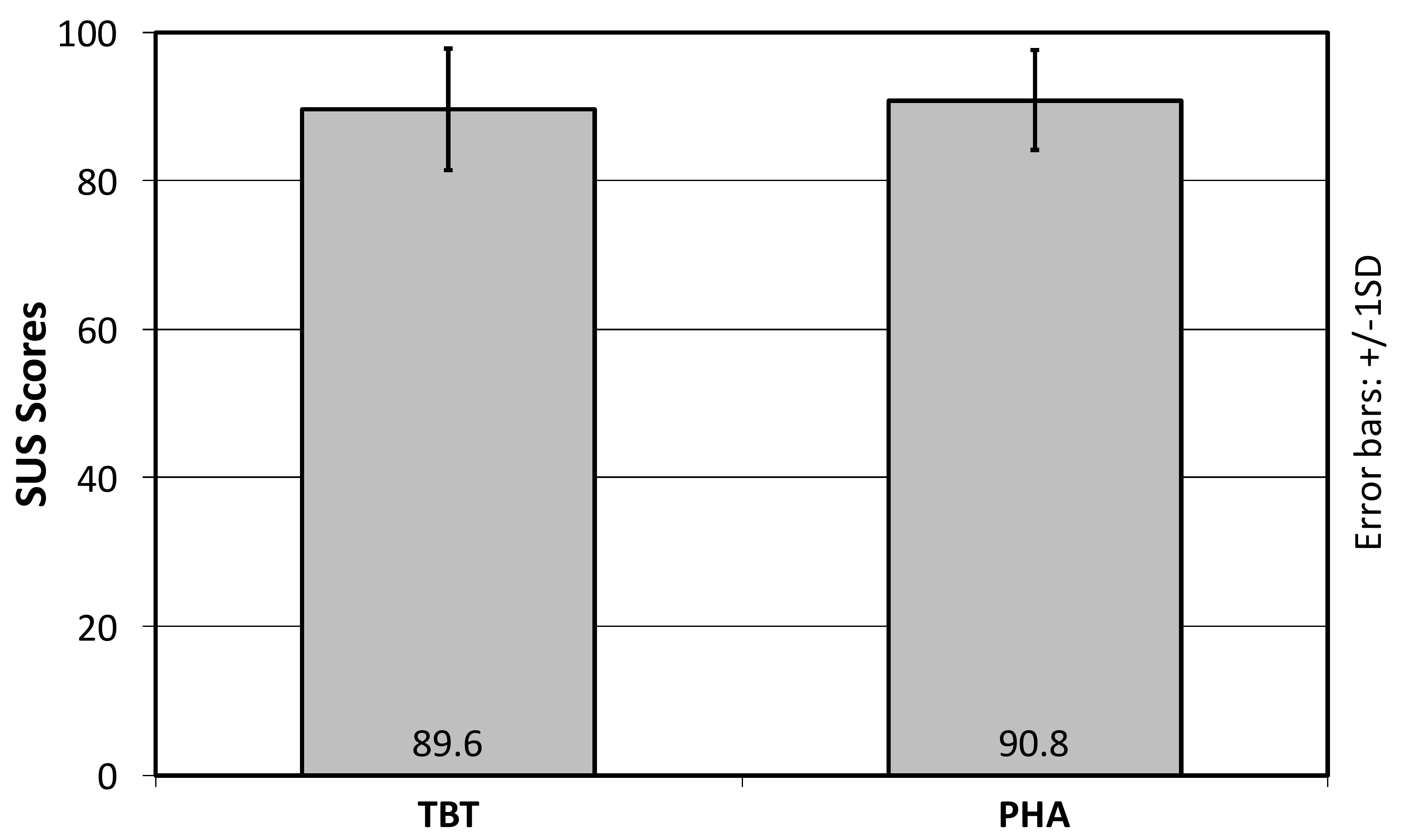}
    \label{fig:results:sus}
  }
  \quad
  \subfloat[]{
    \includegraphics[width=0.30\linewidth]{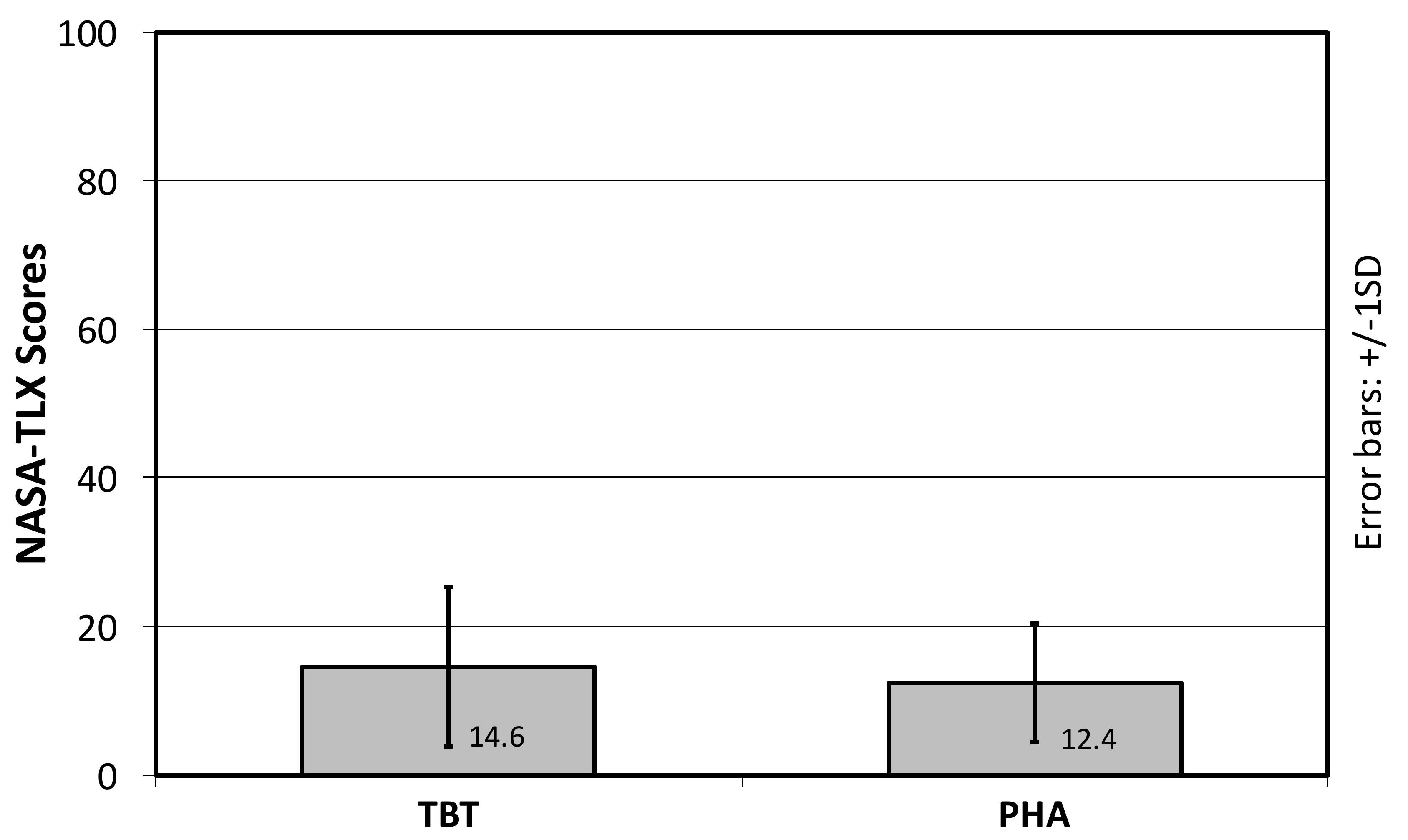}
    \label{fig:results:nasatlx}
  }
  \quad
  \subfloat[]{
    \includegraphics[width=0.30\linewidth]{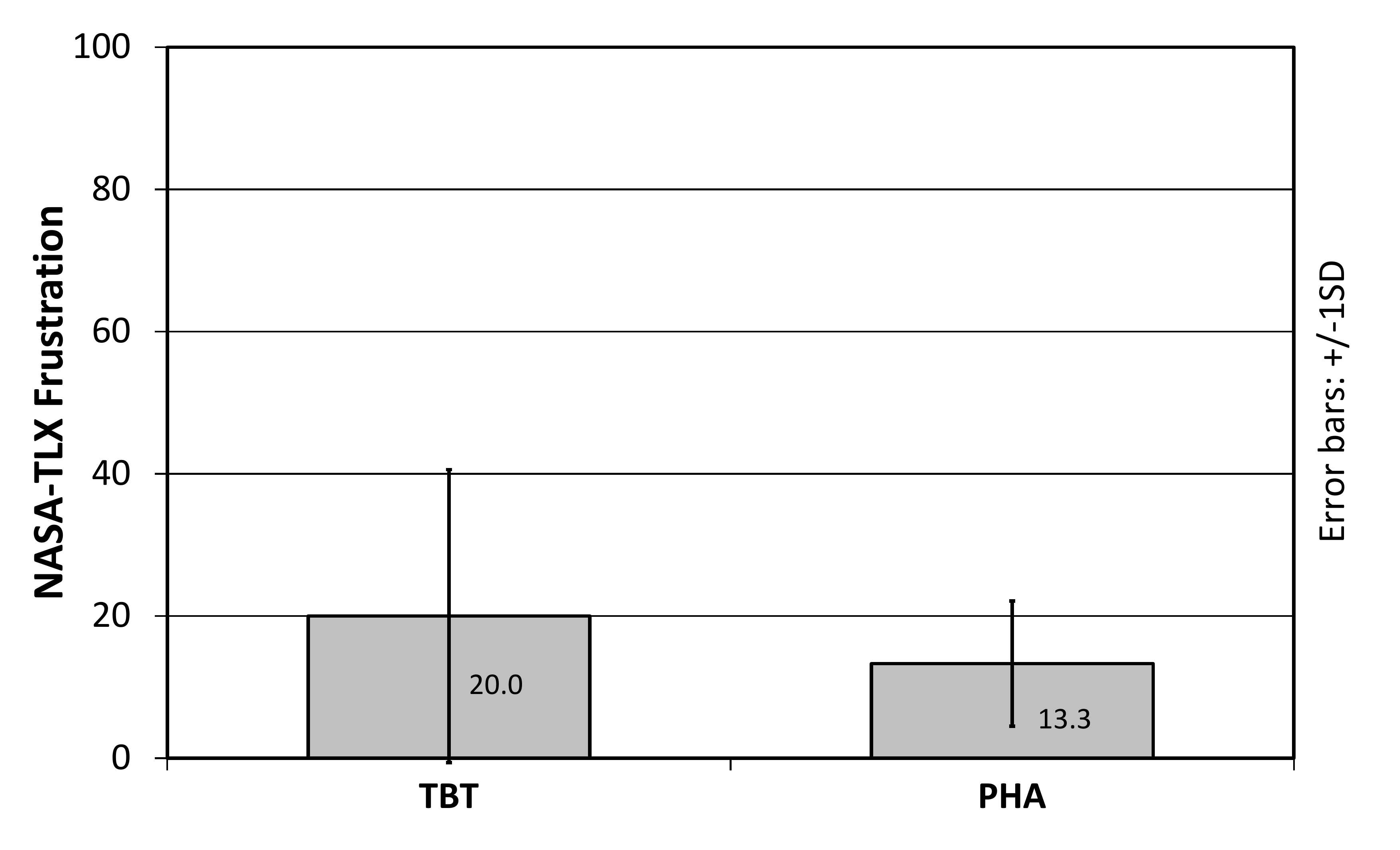}
    \label{fig:results:nasatlxfrust}
  }
  \caption{The results for the baseline (TBT) and Pharos (PHA) conditions for (a) the time the participants needed to complete the routes with both interfaces, (b) how often they looked at the smartwatch, and (c) how confident they felt during navigation. The second row shows the results from (d) the System Usability Scale, (e) the overall NASA-TLX Score, and (f) the results for the NASA-TLX sub-scale of frustration.}
  \label{fig:results}
\end{figure*}

After each condition, participants were asked to draw a cognitive map of the route they were instructed to follow. In the PHA condition, they were also asked to include the location of the TV tower. The concept of cognitive maps was first introduced by Tolman~\cite{tolman1948cognitive} and later adapted and extended to the domain of spatial computing, where cognitive maps are a ``mental representation of people's perception of the real world''~\cite{giannopoulos_influence_2013,garling1984cognitive}. They provide a representation of the spatial knowledge of a user~\cite{montello1996modeling}. The goal was to measure if the users gained more spatial knowledge in the PHA condition.

\begin{figure*}[tb]
  \centering
  \includegraphics[width=\linewidth]{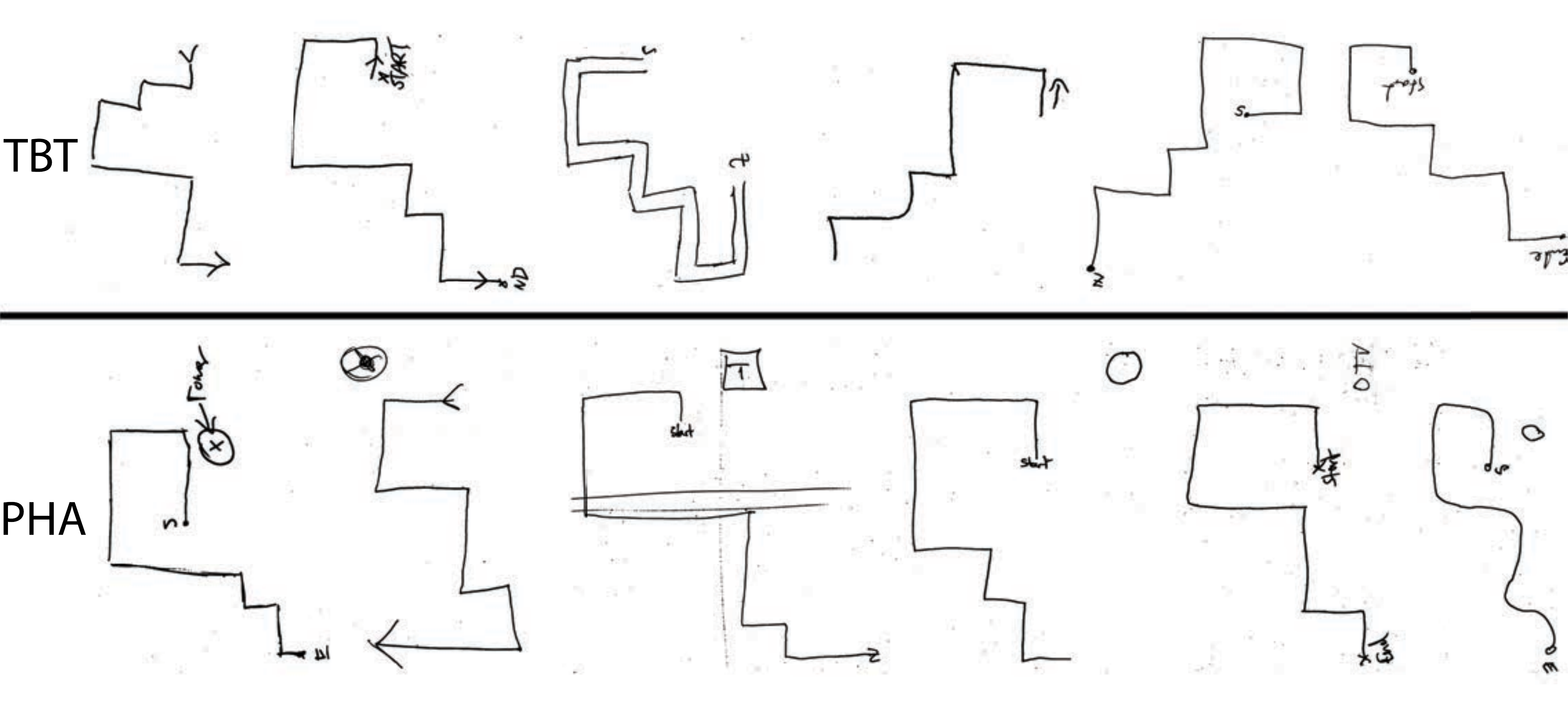}
  \caption{Mental map route sketches for route 2 from all participants in the baseline (TBT) and Pharos (PHA) conditions (rotated and scaled for comparability). First row TBT and second row PHA condition.}
  \label{fig:cogmaps}
\end{figure*}

We also used the NASA-TLX~\cite{Hart1988Development} to measure the perceived workload and the System Usability Scale (SUS)~\cite{Brooke1996SUS} to measure the perceived usability. All questionnaires were filled out for both conditions after the participants had drawn the cognitive map. The total time taken by each participant for the whole study was about 60 minutes. Participants were encouraged to think aloud and to ask questions if necessary. Noteworthy incidents were recorded in writing. A semi-structured interview was conducted with each of the participants after finishing both routes.

\subsection{Results \& Analysis}
All participants were able to complete all the tasks. Figure~\ref{fig:results} summarizes the results of the user study.

On average, the participants took 14 minutes and 4 seconds per route (Route 1: $14{:}39\,m$, Route 2: $13{:}29\,m$). The fastest participant needed 11 minutes and 15 seconds while the slowest one needed 17 minutes and 50 seconds. Participants made a maximum of two errors per route. Most of the errors happened on route 1 ($M{=}0.92$) due to a missing sidewalk at one turning point and only one participant made an error on route 2 ($M{=}0.08$). The participants were slightly faster in the PHA condition compared to TBT (TBT: $M{=}14{:}32\,m$, $SD{=}01{:}21\,m$; PHA: $M{=}13{:}36\,m$, $SD{=}01{:}43\,m$), while they committed almost the same number of errors in both conditions (TBT: $M{=}0.5$, $SD{=}0.65$; PHA: $M{=}0.5$, $SD{=}0.5$). Statistical analysis did not reveal significant differences in either the speed or error rate measures.

Regarding the perceived usability, the SUS scores were high for both the TBT condition ($M{=}89.6$, $SD{=}8.2$) and the PHA condition ($M{=}90.8$, $SD{=}6.8$), see Figure~\ref{fig:results:sus}. That means that the participants had no serious usability problems in both conditions influencing the results. Regarding the perceived taskload, the NASA-TLX values are low and almost the same for both the TBT condition ($M{=}14.6$, $SD{=}10.7$) and the PHA condition ($M{=}12.4$, $SD{=}8.0$), the sub-scale of frustration (``How insecure, discouraged, irritated, stressed, and annoyed were you?'') differs (TBT: $M{=}20.0$, $SD{=}20.6$; PHA: $M{=}13.3$, $SD{=}8.8$), see Figures~\ref{fig:results:nasatlx} and \ref{fig:results:nasatlxfrust}. This means that both conditions evoke a low workload with a slightly higher frustration for the TBT condition. However, statistical analysis did neither reveal a significant difference for the sub-scale of frustration, nor for the NASA-TLX overall values and the SUS scores.

The confidence ratings resulted in three values per participant per route. On average, the confidence was higher in the Pharos condition ($M{=}6.9$, $SD{=}0.1$) than in the TBT condition ($M{=}6.1$, $SD{=}0.4$), see Figure~\ref{fig:results:conf}. A paired t-Test revealed that the difference is statistically significant ($t(11){=}-5.370, p{<} 0.001$) with a large effect size $r{=}0.81$ (Cohen's $d{=}7.243$). The reduced confidence in the TBT instructions is also apparent in the number of looks at the navigation instructions on the smartwatch. The number of looks is substantially lower for the PHA condition ($M{=}19.6$, $SD{=}8.6$) than for the TBT condition ($M{=}27.6$, $SD{=}11.6$), see Figure~\ref{fig:results:looks}. A paired t-test revealed a statistically significant difference between the conditions ($t(11){=}2.339, p{=}0.039$), with a medium effect size $r{=}0.36$ (Cohen's $d{=}0.74$). In other words, with the Pharos approach of integrating global landmarks into the navigation instructions, participants were not only more confident that they are on the correct route, they also looked at their smartwatch less often than was the case with baseline turn-by-turn instructions. We note that looking at one's smartphone is often substantially more effortful than doing so using a smartwatch, so we expect that the benefits of Pharos may be greater in this respect for smartphone-based navigation.

To assess and analyze the representation of the spatial knowledge of a user while performing the task we used cognitive maps~\cite{montello1996modeling}. The cognitive maps drawn by the participants after each condition were digitized using QGIS\footnote{\url{http://www.qgis.org}}. We performed a bi-dimensional regression following the procedure as described by Friedman~\cite{friedman2003bidimensional}. In general, bi-dimensional regression~\cite{tobler1994bidimensional} ``requires an equal number of points between the configurations to be related'' \cite{giannopoulos_influence_2013}. However, this was not true across participants' sketches of the whole route. Therefore, we performed a bi-dimensional regression only on the turning points, excluding all of the segments between the turning points. A Wilcoxon signed-rank test on the retrieved correlation values of the TBT ($M{=}0.51, STD{=}0.17$) and PHA condition ($M{=}0.63, STD{=}0.11$) showed a significant difference between these two conditions ($Z=-3.04, p{=}0.043$). That means that participants could better remember the route with the Pharos approach than with turn-by-turn instructions. Exemplarily, Figure \ref{fig:cogmaps} shows the sketches drawn by the participants for route 2 in the TBT and PHA condition.

\subsubsection{Qualitative Feedback}
In addition to observations, after walking both routes and testing both variants of instructions, we conducted a semi-structured interview with the participants. Overall they were very satisfied with smartwatch navigation and felt very safe: ``I do not have to remember the way, better than with a smartphone [...]'' as navigation instructions on a smartphone have to be constantly checked (P6). One participant stated that ``smartwatch navigation is really cool'' (P7).

When asking the participants which of the two prototypes they prefer, the answers were mixed, even though PHA outperformed TBT in terms of confidence and spatial knowledge. Seven of them preferred the PHA condition and five the TBT instructions. Participants who preferred the TBT condition thought that the additional information is ``unnecessary'' (P9, P10), because they have ``more to read''. However, the participants thought that Pharos is extremely helpful especially at the starting point, whether they preferred PHA in general (P4, P11) or TBT (P10).

\section{Discussion \& Limitations}
In this paper, we present a novel way to include \emph{global landmarks} in pedestrian navigation instructions, which makes landmark-based navigation much more feasible than approaches using local landmarks. From a technical point of view, we have shown that the visibility of global landmarks can be determined automatically from existing and public available geotagged image content. Our approach of using a convolutional neural network combined with a sliding window is robust and easy to train for new landmarks, as it requires only around 100 images for each landmark. In addition, we presented a complete pipeline to not only determine the visibility of global landmarks, but also to include them in the navigation instructions. It allows to compute the visibility of landmarks beforehand on map and navigation servers so that no further computation on a mobile or wearable device is needed.

From the user's perspective, Pharos offers a lightweight but effective navigation support. In a user study, the Pharos approach outperformed current-state-of-the-art turn-by-turn instructions in important navigation metrics. We showed that small textual changes to the navigation instructions including hints on the location of global landmarks led to significantly more confident users. Additionally, although the Pharos navigation instructions contain more information than traditional turn-by-turn instructions, the participants looked less often at their smartwatch while navigating. We also saw that these changes resulted in users building a better spatial knowledge of their environment. Regarding time and error, we did not find any differences. This is not surprising, as a perfectly working turn-by-turn-based navigation system --- also in the baseline condition the navigation instructions were manually triggered with a very high accuracy current state-of-the-art pedestrian system do not provide (see \nameref{sec:userstudy}) --- is probably impossible to outperform in these metrics. However, we are convinced that there is a need for navigation systems providing additional values, as Pharos does, but without resulting in slower navigation and more errors.

Additionally, all participants of the user study lived in the city the study took place, but were not familiar with most parts of the testing sites. As such, our study suggests that Pharos can be helpful when traveling to unknown places. This is particularly true in places where traditional navigation techniques break down (e.g. the street signs are written in a different language or with different characters).

The user study has also highlighted the benefits of smartwatches for pedestrian navigation in general as already pointed out by related work~\cite{Wenig2015StripeMaps, Wenig:2016:SBI:2935334.2935373}. The participants liked navigating via smartwatch and positively mentioned that, when using a smartwatch, they have their hands free for other interactions.  Additionally, the study has also revealed current problems of pedestrian navigation systems for smartwatches that could be overcome by using Pharos. Due to relatively low position accuracy, they constantly provide the user with information about the next decision point, which leads to low confidence and might prevent the user to build up spatial knowledge of their environment.

We evaluated the Pharos approach with global landmarks using large buildings. However, mountains or downtown skylines are global landmarks and can also be included in navigation instructions. This might be more difficult for two reasons: First, skylines and mountains usually look different from different view angles, which is often not the case for buildings. This results in larger training sets that are needed to determine their visibility. Furthermore, in rural areas the availability of geotagged images (e.g. by services like GSV) to determine the visibility of the landmark could be limited as well. To overcome this, we could extend the pipeline to include approaches that compute the visibility of landmarks in rural areas with the help of digital elevation models (DEMs)~\cite{lee1994visibility} or to use geotagged data from crowd-sourced approaches such as Open Street View\footnote{\url{http://openstreetview.org/}}.

%
%
\section{Conclusion \& Future Work}
This paper demonstrates that the Pharos approach is both (1)~feasible and (2)~and has benefits compared to traditional turn-by-turn (i.e. no landmark) instructions with a user study. This means that global landmarks can be included in the navigation instructions within cities for many routes, and that global landmark-enriched instructions can be necessarily cognizant of when the landmark is visible to a user and when it is not.

 Although we evaluated Pharos for pedestrian navigation in urban areas, we are convinced that it can also be useful in other contexts (e.g. outdoors while hiking) or in more rural areas using mountains as global landmarks. For the future, we also want to exploit the use of Pharos in rural environments. Besides that, we are interested to explore whether Pharos could be applied to other navigation domains, such as biking or driving.

To achieve this we will further extend our pipeline to cover a larger set of global landmarks and also include other global landmarks that make sense for other modalities. Furthermore, the visibility maps could also be used to calculate scenic routes (e.g. for tourists) that guide the users through areas where global landmarks are very often visible.

%
%
\section{Acknowledgments}
This work is supported by the Volkswagen Foundation through a Lichtenberg professorship. 

\emph{Note: This version of the paper contains a fix for a reference issue that appeared in the original version.}

\balance{}

\bibliographystyle{SIGCHI-Reference-Format}
\bibliography{Pharos}

\end{document}